\documentclass[a4paper,11pt]{article}
\usepackage{pos,bm}

\title{Quarkonium Polarization Kinetic Equation from Open Quantum Systems and Effective Field Theories}

\author*[a,b]{Di-Lun Yang}
\author[c]{Xiaojun Yao}

\affiliation[a]{Institute of Physics, Academia Sinica,\\
  11529, Taipei, Taiwan}

\affiliation[b]{Physics Division, National Center for Theoretical Sciences,\\
106319, Taipei, Taiwan}

\affiliation[c]{InQubator for Quantum Simulation, Department of Physics, University of Washington,\\
Seattle, WA 98195, USA}
	
\emailAdd{dilunyang@gmail.com}
\emailAdd{xjyao@uw.edu}

\abstract{Recent measurements of polarization phenomena in relativistic heavy ion collisions have aroused a great interest in understanding dynamical spin evolution of the QCD matter. In particular, the spin alignment signature of $J/\psi$ has been recently observed in Pb-Pb collisions at LHC, which may infer nontrivial spin transport of quarkonia in quark gluon plasmas. Motivated by this, we study the spin-dependent in-medium dynamics of quarkonia by using the potential nonrelativistic QCD (pNRQCD) and the open quantum system framework. By applying the Markovian approximation and Wigner transformation, we systematically derive the Boltzmann transport equation for vector quarkonia with polarization dependence in the quantum optical limit. As opposed to the previous study for the spin-independent case where the collision terms depend on chromoelectric correlators, the new kinetic equation incorporates gauge invariant correlators of chromomagnetic fields that determine the recombination and dissociation terms with polarization dependence at the order we are working in the multipole expansion. In the quantum Brownian motion limit, the Lindblad equation with new transport coefficients defined in terms of the chromomagnetic field correlators have also been derived. Our formalism is generic and valid for both weakly-coupled and strongly-coupled quark gluon plasmas. It may be further applied to study spin alignment of vector quarkonia in heavy ion collisions.}

\FullConference{The XVIth Quark Confinement and the Hadron Spectrum Conference (QCHSC24)\\
 19-24 August, 2024\\
 Cairns Convention Centre, Cairns, Queensland, Australia\\}

\begin{document}
\maketitle

\section{Introduction}
Heavy quarkonia have long been recognized as valuable probes of the quark-gluon plasma (QGP) in heavy ion collisions, initially for their static plasma screening effects~\cite{Matsui:1986dk,Karsch:1987pv}. Recent insights highlight dynamic medium effects, including dissociation~\cite{Laine:2006ns,Beraudo:2007ky,Brambilla:2008cx} and recombination~\cite{Thews:2000rj}, which influence quarkonium transport. Extensive studies have developed semiclassical transport equations that account for these effects. See Ref.~\cite{Andronic:2024oxz} for a comprehensive overview of different approaches and more references therein. The open quantum system approach is now a modern framework for analyzing heavy quark-antiquark pairs interacting with a thermal environment ~\cite{Rothkopf:2019ipj,Akamatsu:2020ypb,Yao:2021lus,Sharma:2021vvu}, which can be further integrated with effective field theories (EFTs) like potential nonrelativistic QCD (pNRQCD)~\cite{Brambilla:1999xf,Brambilla:2004jw,Fleming:2005pd}. This integration facilitates the derivation of both semiclassical Boltzmann and Lindblad equations to study quarkonium dynamics. The underlying assumption of the theoretical frameworks is based on the sufficiently weak interaction between the quarknoium as a subsystem and the QGP as an environment due to the separation of scales. 

On the other hand, unlike the previous studies focusing on the observable pertinent to the energy-momentum transport such as the nuclear modification factor of quarkonia, new observable associated with the quarkonium polarization may further reveal the spin transport of heavy quarks in heavy ion collisions. 
Recent experimental results show significant $J/\psi$ spin alignment with respect to the event plane in Pb-Pb collisions measured by the ALICE Collaboration~\cite{ALICE:2022dyy}. See also Ref.~\cite{Cheung:2022nnq,Zhao:2023plc,Chen:2025mrf,Liang:2025hxw} for some recent theoretical studies. In fact, there have been previous observations of the spin alignment for light mesons in heavy ion collisions by both RHIC and LHC~\cite{ALICE:2019aid,STAR:2022fan}, whereas the results have nontrivial flavor and collision-energy dependence. This suggests different underlying mechanisms influenced by fluctuating sources, affecting light (strange) vector meson spin alignment compared to weak global spin polarization in $\Lambda$ hyperons~\cite{STAR:2019erd}. Recent studies using quantum kinetic theory for spin transport, constructed from the Wigner-function approach~\cite{Chen:2012ca,Hidaka:2016yjf,Hattori:2019ahi,Yang:2020hri,Hidaka:2022dmn} , have explored spin alignment due to chromoelectromagnetic fields in the QGP or pre-equilibrium phase~\cite{Luo:2021uog,Muller:2021hpe,Yang:2021fea}, indicating that gluonic fluctuations are critical for understanding vector meson spin alignment in heavy ion collisions~\cite{Kumar:2022ylt,Kumar:2023ghs,Yang:2024qpy}.

However, most existing theoretical studies have focused on light vector mesons, treating quarks and antiquarks' spin dynamics independently. This approach may be inadequate for quarkonium, a bound state, particularly in low-temperature QGP phases. In a recent study \cite{Yang:2024ejk}, a first-principle derivation of the transport equation with polarization dependence for vector quarknia, allowing for a comprehensive study of time evolution of polarization in $S$-wave spin-triplet quarkonia and incorporating nonperturbative effects through gauge-invariant chromomagnetic correlators, was achieved by combining the open quantum system framework and pNRQCD as the extension of previous studies that are independent of the quarkonium spin \cite{Yao:2018nmy,Yao:2020eqy}. In this manuscript, we will review the major findings of Ref.~\cite{Yang:2024ejk}. For brevity, only the key steps, assumptions, and results are presented here and most technical details are omitted.

The manuscript is organized as follows: Sec.~\ref{sec:theory_framework} reviews the open quantum system and pNRQCD as the adopted theoretical frameworks, while Sec.~\ref{sec:OQS_to_KE} explains how we derive the spin-dependent Boltzmann equation for vector quarknoia. In Sec.~\ref{sec:discussions}, we discuss the potential phenomenological and future directions. Throughout this manuscript, we adopt the mostly minus Minkowski metric signature, use Greek and roman indices for spacetime and spatial components, and represent spatial Euclidean vectors in bold. Dot products of spatial vectors are denoted as ${\bm A}\cdot {\bm B} = A_i B_i$. We also use $[,]$ and $\{,\}$ to represent the commutation and anti-commutation relations, respectively.
\section{Theoretical frameworks}\label{sec:theory_framework}
We begin with a brief review of the framework of open quantum system (OQS). The general idea of OQS is to track the dynamical evolution of the density matrix of a subsystem immersed in an environment. Considering the full Hamiltonian with the decomposition,
\begin{eqnarray}
H=H_S+H_E+H_I \,,
\end{eqnarray}
where the subsystem Hamiltonian $H_S$ evolving inside an environment described by the Hamiltonian $H_E$ and
we further assume the interaction Hamiltonian can be written as
\begin{eqnarray}	
\label{eqn:H_I}
H_I = \sum_\alpha O^{(S)}_\alpha \otimes O^{(E)}_\alpha \,
\end{eqnarray}
with $O^{(S)}_\alpha$ and $O^{(E)}_\alpha$ being Hermitian operators that only act on the subsystem and the environment, respectively. From the Schr\"odinger equation, the time evolution of the density matrix for the full system is governed by
\begin{eqnarray}\label{eq:Schrodinger_eq}
\frac{d\rho^{(\rm int)}(t)}{dt}=-i[H^{(\rm int)}_I,\rho^{(\rm int)}(t)]
\end{eqnarray} 
in the interaction picture. We further assume that the total density matrix is factorized at time $t_i$, based on the weak coupling between the subsystem and the environment, and the environment density matrix is a thermal (Gibbs) state that is only time dependent through its temperature
\begin{align}
	\label{eqn:rho_factorize}
	\rho^{(\rm int)}(t_i) = \rho_S^{({\rm int})}(t_i) \otimes \rho_E^{({\rm int})}(t_i) \,,\quad \rho_E^{({\rm int})}(t_i) = \frac{e^{-\beta(t_i) H_E}}{Z_E} = \frac{e^{-\beta(t_i) H_E}}{{\rm Tr}_E(e^{-\beta(t_i) H_E}) } \equiv \rho_T(t_i) \,,
\end{align}
where $\beta=\frac{1}{T}$ is given by the inverse of the environment temperature.
In order to capture the dissipation of the subsystem as a non-unitary process,  we then trace out the environment degrees of freedom and obtain the reduced density matrix of the subsystem $\rho_S^{({\rm int})}(t) = {\rm Tr}_E [\rho^{(\rm int)}(t)] $. The perturbative solution of Eq.~(\ref{eq:Schrodinger_eq}) for the traced density matrix with $t_i=0$ is accordingly given by
\begin{align}
	\label{eqn:pre_lindblad}
	&\rho_S^{({\rm int})}(t) = \rho_S^{({\rm int})}(0) - \int_0^t {\rm d}t_1 \int_0^t {\rm d}t_2 \frac{{\rm sign}(t_1-t_2)}{2} D_{\alpha\beta}(t_1,t_2) \left[ O^{(S)}_\alpha(t_1) O^{(S)}_\beta(t_2), \rho_S^{({\rm int})}(0) \right] \nonumber\\
	& + \int_0^t {\rm d}t_1 \int_0^t {\rm d}t_2  D_{\alpha\beta}(t_1,t_2) \left( O^{(S)}_\beta(t_2) \rho_S^{({\rm int})}(0) O^{(S)}_\alpha(t_1) - \frac{1}{2}\left\{ O^{(S)}_\alpha(t_1) O^{(S)}_\beta(t_2), \rho_S^{({\rm int})}(0)\right\} \right),
\end{align}
where $\alpha,\beta$ are implicitly summed over and environment correlation function is defined as
\begin{align}
	D_{\alpha\beta}(t_1,t_2) = {\rm Tr}_E\left( O^{(E)}_\alpha(t_1) O^{(E)}_\beta(t_2) \rho_T(0) \right) \,,
\end{align}
which characterizes the interaction between the subsystem and environment through $H_{\rm I}$. Note that the master equation in Eq.~(\ref{eqn:pre_lindblad}) is a finite difference equation, which requires further approximation to arrive at a kinetic equation as will be shown in the next section. 

To proceed with the application for quarkonium transport in heavy ion collisions, we need to introduce an explicit form of Hamiltonian with the underlying quantum field theory. For non-relativistic quarkonia, we may introduce the following scales with a hierarchy, $M\gg Mv\gg Mv^2,\,\Lambda_{QCD}$, where $M$ and $v$ represent the heavy-quark mass and the relative velocity of the heavy quark $Q$ and anti heavy quark $\bar{Q}$ pair, respectively. Here $Mv$ and $Mv^2$ correspond to the soft and ultra-soft scales that characterize the inverse of a typical relative distance between the $Q\bar{Q}$ pair and the scale of binding energy, respectively. At the typical energy scale smaller than the soft scale, pNRQCD serves as an effective description for non-relativistic quarkonia, where the building block is a heavy quark-antiquark composite field
\begin{eqnarray}
\Phi_{ij}^{s_1s_2}({\bm x}_1, {\bm x}_2, t) = \psi^{s_1}_{i}({\bm x}_1, t) \otimes \chi^{\dagger s_2}_{j}({\bm x}_2, t)\,,
\end{eqnarray}
which annihilates a pair of heavy quark and antiquark at position ${\bm x}_1$ and ${\bm x}_2$, respectively, through the heavy quark (antiquark) annihilation operator, $\psi^{s_1}_{i}$ ($\chi^{\dagger s_2}_{j}$). Here the superscripts $s_{1,2}$ and subscripts $i,j$ of $\Phi_{ij}^{s_1s_2}({\bm x}_1, {\bm x}_2, t)$ represent the spin and color indices. Then we decompose the composite field as a color-singlet field $S$ and a color-octet field $O^a$ with $a=1,2,\dots ,N_c^2-1$,
\begin{align}
	\label{eqn:phi_ij}
	\Phi_{ij}({\boldsymbol x}_1, {\boldsymbol x}_2, t) = U_{ik}({\bm x}_1, {\bm R}) \left( \frac{\delta_{k\ell}}{\sqrt{N_c}} S({\bm R},{\bm r},t) + \frac{T^a_{k\ell}}{\sqrt{T_F}} O^a({\bm R},{\bm r},t) \right) U_{\ell j}({\bm R}, {\bm x}_2)\,,
\end{align}
where ${\bm R}=({\bm x}_1+{\bm x}_2)/2$ and ${\bm r}={\bm x}_1-{\bm x}_2$ are the center-of-mass (cm) and relative positions, respectively, and 
\begin{align}
	U({\bm x}, {\bm y}) \equiv \mathcal{P}\exp\left( ig\int_{\bm y}^{\bm x} {\bm A}({\bm z}) \cdot d{\bm z} \right) \,,
\end{align}
denotes a spatial Wilson line with a straight path from $\bm y$ to $\bm x$. For brevity, we omitted the spin indices above. $\mathcal{P}$ denotes path ordering. Gauge fields are matrix variables, i.e., $A_i = A_i^aT^a$, where $T^a$ is the generator of the SU($N_c$) group and is normalized as ${\rm Tr}_c(T^aT^b) = T_F\delta^{ab}$ with $T_F=1/2$. Also, we only consider the $S$-wave composite field and further make the spin decomposition as
\begin{eqnarray}
	S=\frac{1}{\sqrt{2}}\left(IS_1+\sum_{\lambda=0,\pm 1}S_{\lambda i}\sigma_{i}\right),\quad
	O^a=\frac{1}{\sqrt{2}}\left(IO^a_1+\sum_{\lambda=0,\pm 1}O^a_{\lambda i}\sigma_i\right),
\end{eqnarray}
where $S_1, O^a_1$ and $S_{\lambda i}, O^a_{\lambda i}$ correspond to the spin-singlet and spin-triplet components, respectively. Physically, we may regard the spin-singlet and spin-triplet components as the pseudo-scalar and vector quarkonium fields. Eventually, the relevant interaction terms, characterizing the transitions between color-singlet and color-octet fields and between spin-singlet and spin-triplet fields, in the effective Lagrangian, up to $\mathcal{O}(|\bm r|)$ in terms of the multipole expansion and to $\mathcal{O}(M^{-1})$ in the nonrelativistic expansion, is given by
\begin{align}\nonumber\label{eq:int_L}
	\mathcal{L}_{I}&=V_A \sqrt{\frac{T_F}{N_c}} \bigg( S_{1}^\dagger({\boldsymbol R}, {\boldsymbol r},t) {\bm r}\cdot g{\bm E}^a({\bm R},t) O^{a}_{1}({\boldsymbol R}, {\boldsymbol r},t)+ \!\sum_{\lambda}\!\! S_{\lambda i}^\dagger({\boldsymbol R}, {\boldsymbol r},t) {\bm r}\cdot g{\bm E}^a({\bm R},t) O^{a}_{\lambda i}({\boldsymbol R}, {\boldsymbol r},t)+{\rm h.c.}\bigg)
	\\\nonumber
	&+\frac{c_4}{M} V_{A}^s \sqrt{\frac{T_F}{N_c}} \sum_\lambda \left[ S^\dagger_1({\bm R},{\bm r},t) gB_i^a({\boldsymbol R},t) O^a_{\lambda i}({\bm R},{\bm r},t) + \, S^\dagger_{\lambda i}({\bm R},{\bm r},t) gB_i^a({\boldsymbol R},t) O^a_1({\bm R},{\bm r},t)+{\rm h.c.} \right],
	\\
\end{align} 
where ${\bm E}^a$ and ${\bm B}^a$ denote the chromoelectric and chromomagnetic fields, respectively. The involved coefficients $V_{A}$, $V_{A}^s$, and $c_4$ need to be derived via the matching with NRQCD~\cite{Bodwin:1994jh}. Note that the chromomagnetic terms are essential for the spin-dependent transitions. From $\mathcal{L}_{I}$, one can read out $O^{(S)}_\alpha$ and $O^{(E)}_\alpha$ and derive the kinetic equation from Eq.~(\ref{eqn:pre_lindblad}).  

\section{Spin-dependent Boltzmann equation for vector quarknoia}\label{sec:OQS_to_KE} 
To work out the kinetic equation from OQS, we will consider the quantum optical limit such that the system relaxation time $\tau_{R}$ is much larger than both the environment correlation time $\tau_{E}$ and the intrinsic timescale of the system $\tau_{S}$, i.e., $\tau_{R}\gg \tau_{E}, \tau_{S}$. For the quarkonium transport in QGP, we have $\tau_{S}\sim (Mv^2)^{-1}$ as the inverse of the binding energy and $\tau_{E}\sim T^{-1}$ with $T$ being the temperature of QGP. Under the hierarchy of energy scales $M\gg Mv \gg Mv^2,T,\Lambda_{\rm QCD}$ for the low-temperature region, the multipole expansion is justified $|\bm r|T\sim (Mv)^{-1}T\ll 1$. In addition, we will adopt the Markovian approximation by the coarse graining such that $\tau_{R}\gg t\gg\tau_{E}$ for $t$ being the relevant time scale of the kinetic equation. 

Given the binding energy is much larger than the dissociation/recombination rates from $\tau_{R}\gg\tau_{S}$, we can introduce a complete set from eigenstates of the subsystem. From Eq.~(\ref{eqn:pre_lindblad}), we may recast the finite difference equation in the Schr\"odinger picture as 
\begin{eqnarray}
\rho_S(t) = \rho_S(0) -it[H_S,\rho_S(0)] -i\sum_{a,b} \sigma_{ab}(t) [L_{ab}, \rho_S(0) ]+\tilde{C}^{+}(t)-\tilde{C}^{-}(t), 
\label{eqn:lindblad}
\end{eqnarray}
where $L_{ab} \equiv |E_a\rangle \langle E_b|$ and $\{ |E_a\rangle \}$ consists of eigenstates of $H_S$, i.e., $H_S|E_a\rangle = E_a|E_a\rangle$ and forms a complete set of states in the Hilbert space of the subsystem. Here $\tilde{C}^{+}(t)$ and $\tilde{C}^{-}(t)$ are responsible for the recombination and dissociation effects, respectively, which are given by
\begin{eqnarray}
\tilde{C}^{+}(t)\equiv  \sum_{a,b,c,d} \gamma_{ab,cd} (t) L_{ab} \rho_S(0) L^{\dagger}_{cd},
\quad
\tilde{C}^{-}(t)\equiv  \sum_{a,b,c,d} \gamma_{ab,cd} (t) \frac{1}{2} \{ L^{\dagger}_{cd}L_{ab}, \rho_S(0)\}, 
\end{eqnarray}
where explicit expressions of $\sigma_{ab}(t)$ and $\gamma_{ab,cd} (t)$ depend on the detailed interactions governed by the underlying quantum field theory. That is, pNRQCD for our interest. From the Markovian approximation, one can show $t\gg\tau_{E}$ yields $\tilde{C}^{\pm}(t)\propto t$, which enable us to recast Eq.~(\ref{eqn:lindblad}) into a differential equation by taking $\tau_{R}\gg t\rightarrow 0$. Finally, we will conduct the Wigner transformation to rewrite the expectation values of the subsystem density matrix, sandwiched by states, in terms of the distribution functions for quarkonia in phase space. For example, for the color-singlet and spin-triplet states, we may introduce 
\begin{eqnarray}
f_{\lambda}({\bm x}, {\bm k}, t) \equiv \int\frac{d^3k'}{(2\pi)^3} e^{i {\bm k}'\cdot {\bm x} } \Big\langle  {\bm k}+\frac{{\bm k}'}{2}, \lambda  \Big| \rho_S(t)  \Big|   {\bm k}-\frac{{\bm k}'}{2} , \lambda \Big\rangle \,,
\end{eqnarray} 
as the distribution function of a vector quarkonium in phase space. For simplicity, we only consider the radial ground states such as $J/\psi$ and $\eta_c$ and accordingly drop the additional quantum numbers for the states. Similarly, we may also define the distribution function for the color-octet and spin-singlet $Q\bar{Q}$ pair, $f_{Q\bar{Q}}^{(8)}$, through 
\begin{eqnarray}
&&\int\frac{d^3 k'}{(2\pi)^3} e^{i{\bm k}'\cdot{\bm x}} \Big\langle {\bm k}+\frac{{\bm k}'}{2}, {\bm p}_{1\rm{rel}},a \Big| \rho^{(8)}_S(t) \Big| {\bm k}-\frac{{\bm k}'}{2}, {\bm p}_{2\rm{rel}} ,a'\Big\rangle \nonumber\\
&&\approx (2\pi)^3 \delta^3({\bm p}_{1\rm{rel}} - {\bm p}_{2\rm{rel}}) \delta^{aa'}f_{Q\bar{Q}}^{(8)}({\bm x}, {\bm k}, {\bm x}_0, {\bm p}_{\rm{rel}} ,t ) \,,
\end{eqnarray}
where we have only kept the diagonal term in color space and the leading-order term of the gradient expansion. Note that $f_{Q\bar{Q}}^{(8)}$ not only depends on ${\bm k}$, the center-of-mass momentum, but also ${\bm p}_{\rm{rel}}\equiv({\bm p}_{1\rm{rel}}+{\bm p}_{2\rm{rel}})/2$, the relative one between $Q$ and $\bar{Q}$. By the same manner, one can also define the distribution functions for the color-triplet and spin-triplet states and for the color-singlet and spin-singlet state.

Subsequently, taking a time derivative of Eq.~(\ref{eqn:lindblad})\footnote{This procedure is equivalent to take $\lim_{t\rightarrow 0}(\rho_S(t)- \rho_S(0))/t$. } sandwiched by the color-singlet and spin-triplet states and carrying out the Wigner transformation result in a Boltzmann equation for the vector quarkonium,
\begin{align}
	\label{eqn:Boltzmann_p}
	\frac{\partial}{\partial t} f_{\lambda}({\bm x}, {\bm k}, t) + \frac{{\bm k}}{2M} \cdot \nabla_{\bm x} f_{\lambda}({\bm x}, {\bm k}, t)
	= \mathcal{C}_{\lambda}^+({\bm x}, {\bm k}, t)[f_{Q\bar{Q}}^{(8)}] - \mathcal{C}_{\lambda}^-({\bm x}, {\bm k}, t)[f_{\lambda}]  \,,
\end{align}
where the collision kernels
\begin{align}
	\mathcal{C}_{\lambda}^+({\bm x}, {\bm k}, t)[f_{Q\bar{Q}}^{(8)}] &= \int\frac{d^3 k'}{(2\pi)^3}e^{i {\bm k}'\cdot {\bm x} } \Big\langle {\bm k}+\frac{{\bm k}'}{2},\lambda\Big| \partial_t \widetilde{\mathcal{C}}^{+}(t)\Big|{\bm k}-\frac{{\bm k}'}{2},\lambda\Big\rangle \,, \nonumber\\
	\mathcal{C}_{\lambda}^-({\bm x}, {\bm k}, t)[f_{\lambda}] &= \int\frac{d^3 k'}{(2\pi)^3}e^{i {\bm k}'\cdot {\bm x} } \Big\langle {\bm k}+\frac{{\bm k}'}{2},\lambda\Big| \partial_t \widetilde{\mathcal{C}}^{-}(t)\Big|{\bm k}-\frac{{\bm k}'}{2},\lambda\Big\rangle \,,
\end{align}
explicitly depend on polarization. Note that $\mathcal{C}_{\lambda}^+$ and $\mathcal{C}_{\lambda}^-$ corresponding to the recombination term and to the dissociation term are proportional to the color-octet distribution function as a deconfined state and the color-singlet one as a bound state, respectively. Here the spatial derivative term in the free-streaming part is led by the remaining terms proportional to $[H_S,\rho_S(0)]$ and $\sigma_{ab}$ in Eq.~(\ref{eqn:lindblad})~\cite{Yao:2018nmy}.

Based on the interaction terms in Eq.~(\ref{eq:int_L}) with the canonical quantization of the quarkonium fields (see Ref.~\cite{Yang:2024ejk} for details), the dissociation term explicitly reads
\begin{align}
	\label{eq:C_dissociation}
	\mathcal{C}_{\lambda}^-({\bm x}, {\bm k}, t)[f_{\lambda}] = \int\frac{d^3p_{\rm{cm}}}{(2\pi)^3} \frac{d^3p_{\rm{rel}}}{(2\pi)^3}  \int d^4q \, \delta(E^{\lambda}-E_{p\rm rel}-q_0)
	\delta^3({\bm k}-{\bm p}_{\rm cm}-{\bm q})
	|\mathcal{M}_d|^2 f_{\lambda}({\bm x}, {\bm k}, t) \,,
\end{align} 
where the scattering-amplitude square takes the form 
\begin{align}\label{eq:diss_collisions}
|\mathcal{M}_d|^2 = \frac{V_A^2T_F}{N_c}\tilde{g}^{E++}_{ij}(q)\langle \psi^{\lambda} | r_i | \Psi_{{\bm p}_{\rm{rel}}}^{\lambda} \rangle \langle \Psi_{{\bm p}_{\rm rel}}^{\lambda} | r_j | \psi^{\lambda} \rangle + \frac{(c_4  V_{A}^s)^2T_F}{M^2N_c}\tilde{g}^{B++}_{ij}(q) \varepsilon^*_{\lambda i}\varepsilon_{\lambda j}  	|\langle \psi^{\lambda} |  \Psi_{{\bm p}_{\rm rel}} \rangle|^2 \,,
\end{align}
with $\varepsilon_{\lambda j}$ being the polarization vector of the vector quarkonium.
Here $E^{\lambda}$ and $E_{p\rm rel}$ correspond to the eigenenergies of color-singlet and color-octet states along with their wave functions, $\psi^{\lambda}$ and $\Psi_{{\bm p}_{\rm rel}}^{\lambda}$, solved from the wave equations with static potentials. We have also implicitly assumed the degeneracy of the energies for $\Psi_{{\bm p}_{\rm rel}}^{\lambda}$ and $\Psi_{{\bm p}_{\rm rel}}$. On the other hand, $q_0$ denotes the energy of scattered gluons. Taking charm quarks as example, the dissociation term basically delineates the scattering process, $J/\psi+g\rightarrow c+\bar{c}$. The chromoelectric and chromomagnetic field correlators for the inclusive dissociation process shown in Eq.~(\ref{eq:diss_collisions}) are given by
\begin{align}
	\tilde{g}^{V++}_{ij}(q_0) &= \int{\rm d}t\,e^{iq_0t} g_{ij}^{V++}(t) \,, \nonumber\\
	g_{ij}^{V++}(t) &= {\rm Tr}_E\Big\{ gV_i^a({\bm R},t) W^{ac}[({\bm R},t),({\bm R},\infty)] W^{cb}[({\bm R},\infty),({\bm R},0)] gV_j^b({\bm R},0) \rho_T(0) \Big\} \nonumber\\
	&= {\rm Tr}_E\Big\{ gV_i^a({\bm R},t) W^{ab}[({\bm R},t),({\bm R},0)] gV_j^b({\bm R},0) \rho_T(0) \Big\} \,,
\end{align}
where $V=E$ or $B$ and $W$ represents the adjoint Wilson line for gauge invariance. The ${\bm R}$ dependence in the fields is irrelevant due to the spatial translational invariance. In general, when one considers the momentum dependence in the dissociation process, the Wilson lines are extended from the finite space-time points to the future space-time infinity, which form a staple shape. Technically, the staple shape stems from sophisticated summation of interactions at higher orders in the coupling but at the monopole order in the multipole expansion
presented in Ref.~\cite{Yao:2020eqy}. However, it is physically intuitive that the Wilson lines should end at future infinity because the final state is now a color octet. From Eq.~(\ref{eq:diss_collisions}), it is obvious to see that the chromoelectic correlator does not affect the polarization, while the chromomagnetic correlator is coupled with the polarization vectors and thus modifies the quarkonium polarization. Finally, one may write down the recombination term, 
\begin{align}
	\label{eq:C_recombination}
	&\mathcal{C}_{n\ell,\lambda}^+({\bm x}, {\bm k}, t)[f_{Q\bar{Q}}^{(8)},f^{(8)}_{Q\bar{Q}\lambda}] = \int\frac{d^3p_{\rm{cm}}}{(2\pi)^3} \frac{d^3p_{\rm{rel}}}{(2\pi)^3} \int d^4q \,  \delta(E^\lambda -E_{p_{\rm rel}}+q_0)\delta^3({\bm k}-{\bm p}_{\rm cm}+{\bm q}) \nonumber\\
	&\qquad\qquad \times \left( |\mathcal{M}_{r,e}|^2 f_{Q\bar{Q}\lambda}^{(8)}({\bm x}, {\bm p}_{\rm{cm}}, {\bm x}_0, {\bm p}_{\rm{rel}} ,t) + |\mathcal{M}_{r,b}|^2 f_{Q\bar{Q}}^{(8)}({\bm x}, {\bm p}_{\rm{cm}}, {\bm x}_0, {\bm p}_{\rm{rel}} ,t) \right) \,,
\end{align}
where
\begin{align}
	|\mathcal{M}_{r,e}|^2 &= \frac{V_A^2T_F}{N_c} \tilde{g}^{E--}_{ji}(q)\langle \psi^{\lambda} | r_i | \Psi_{{\bm p}_{\rm rel}}^{\lambda} \rangle \langle \Psi_{{\bm p}_{\rm rel}}^{\lambda} | r_j | \psi^{\lambda} \rangle \,, \nonumber
	\\	|\mathcal{M}_{r,b}|^2 &= \frac{(c_4  V_{A}^s)^2T_F}{M^2N_c}\tilde{g}^{B--}_{ji}(q) \varepsilon^*_{\lambda i}\varepsilon_{\lambda j} |\langle \psi^{\lambda} |  \Psi_{{\bm p}_{\rm rel}} \rangle|^2 \,.
\end{align}  
The chromoelectric and chromomagnetic correlators are similarly embedded with the adjoint Wilson lines for gauge invariance,
\begin{align}
	\label{eqn:pre_g--}
	&\big[ g^{V--}_{ji}(t_2, t_1, {\bm R}_2, {\bm R}_1) \big]^{a_2a_1} \equiv {\rm Tr}_E\Big\{ W^{a_2b}[({\bm R}_1,-\infty),({\bm R}_2,-\infty)]  \nonumber\\
	& \times W^{bc}[({\bm R}_2,-\infty),({\bm R}_2,t_2)] gV_j^c({\bm R}_2,t_2) gV_i^d({\bm R}_1,t_1) W^{da_1}[({\bm R}_1,t_1),({\bm R}_1,-\infty)] \rho_T(0) \Big\}  \,,
\end{align}
while the Wilson lines are now extended to the past space-time infinity since the initial state is a color octet. The dissociation and recombination terms are the primary results of this proceeding, manifesting how chromomagnetic fields can induce the spin polarization for vector quarkonia via the transitions between the pseudo-scalar $Q\bar{Q}$ pair (quarkonium) and vector qurkonium ($Q\bar{Q}$ pair). 
 
\section{Phenomenological applications and future directions}\label{sec:discussions}
In Eq.~(\ref{eqn:Boltzmann_p}), a polarization-dependent Boltzmann equation for the vector quarkonium traversing the QGP medium is derived. However, the input of color-octet distribution function is needed, which has to be solved from a coupled kinetic equation tracking the phase-space evolution of $f_{Q\bar{Q}}^{(8)}$. Such a study was previously carried out for spin-independent bottomonia in Ref.~\cite{Yao:2020xzw}, whereas the spin dependent case is expected to be more involved. Moreover, a similar kinetic equation for $f_{Q\bar{Q}\lambda}^{(8)}$ is not yet available. Even though the chromoelectric fields do not couple to the polarization vectors for both the dissociation and recombination terms as opposed to the chromomagnetic fields that explicitly modify the polarization transport, the electric interaction may still implicitly affect the polarization especially in the presence of a nontrivial $f_{Q\bar{Q}\lambda}^{(8)}$ originating from initial polarization of the $Q\bar{Q}$ pair or far-from-equilibrium dynamics in the initial stage of heavy ion collisions. Integrating the current framework with the quantum kinetic theory for spin transport of quarks under background color fields~\cite{Luo:2021uog,Muller:2021hpe,Yang:2021fea} may be helpful to extract the information of $f_{Q\bar{Q}\lambda}^{(8)}$, which requires future studies. 

One of the potential phenomenological applications of the constructed theoretical framework is to evaluate the spin alignment of vector quarkonia like $J/\psi$. Given $f_{\lambda}$ solved from the kinetic equation
Eq.~(\ref{eqn:Boltzmann_p}), we may compute the spin alignment observable defined by the diagonal components of the normalized spin density matrix, 
\begin{equation}
	\rho_{\lambda\lambda}(\bm k)=\frac{\int {\rm d}\Sigma_{x}\cdot k \,f_{\lambda}({\bm x}, {\bm k}, t)}{\int {\rm d}\Sigma_{x}\cdot k\,\sum_{\lambda'=\pm 1,0}\,f_{\lambda'}({\bm x}, {\bm k}, t)} \,, \label{diag_spindensity}
\end{equation}
where ${\rm d}\Sigma_{x\mu}$ represents an element on a freeze-out hypersurface and $k^\mu=(k^0,{\bm k})$ with $k^0=\sqrt{{\bm k}^2+m_\lambda^2}$ and $m_{\lambda}$ being the mass of the vector meson. Nontrivial spin alignment is then characterized by the deviation of $\rho_{00}$ from $\frac{1}{3}$.
When $f_{0}=f_{+}=f_{-}$, $\rho_{00}=\frac{1}{3}$, there is no spin alignment. On the other hand, the quantity $\rho_{++}-\rho_{--}$ is proportional to the spin polarization of the vector meson.

Nevertheless, there exists a caveat for the application of the current formalism to charmonia stemming from the validity of multipole expansion in pNRQCD. Since $|\bm r|^{-1}\sim Mv\sim 0.9$ GeV for charmonia and $T_{\rm QGP}\sim 0.15-0.45$ GeV, the $|\bm r|T$ expansion may break down in early times of the QGP phase. Considering e.g., spin alignment of $J/\psi$, we may however still apply the formalism to the low-temperature region close to chemical freeze-out, where the recombination effect should play a dominant role. In addition, except for the quantum optical limit, one may consider the quantum Brownian motion limit such that $\tau_{R},\tau_{S}\gg \tau_{E}$, which results in the hierarchy of energy scales $M\gg Mv \gg T\gg Mv^2,\Lambda_{\rm QCD}$ for the high-temperature region.  In such a case, one may derive the Lindblad equation tracking the time evolution of the subsystem density matrix. However, the multipole expansion becomes invalid in the high-temperature limit for $T\gg Mv$, while we may still consider the contribution from chromomgantic fields in the $M^{-1}$ expansion. In such a scenario, similar to the collision terms for the Boltzmann equation in the quantum optical limit, the correlators of chromomagnetic fields yield the spin-flipping effect in the Lindblad equation as well~\cite{Yang:2024ejk}. In principle, such chromomagnetic correlators could be calculated perturbatively at weak coupling or non-perturbatively via the lattice simulations or other methods. Some related studies for the similar computations of chromoelectric correlators have been recently carried out~\cite{Yao:2018sgn,Binder:2021otw,Nijs:2023dks,Nijs:2023dbc,Leino:2024pen,Scheihing-Hitschfeld:2023tuz}. In summary, the combination of OPS and pNRQCD gives rise a useful transport theory for studying polarization phenomena of vector quarkonia through the correlators of chromomagnetic fields,  while further investigation is necessary to understand the spin transport of heavy quarkonia.     

\acknowledgments
D.-L. Y. thanks Yukinao Akamatsu, Nora Brambilla, Thomas Mehen, Joan Soto, and Antonio Vairo for fruitful discussions during the participation of QCHSC24. D.-L. Y. is supported by National Science and Technology Council (Taiwan) under Grants No. MOST 110-2112-M-001-070-MY3 and No. NSTC 113-2628-M-001-009-MY4 and by Academia Sinica under Project No. AS-CDA-114-M01. X. Y. is supported
by the U.S. Department of Energy, Office of Science, Office of Nuclear Physics, InQubator
for Quantum Simulation (IQuS) (https://iqus.uw.edu) under Award No. DOE (NP)
DE-SC0020970 via the program on Quantum Horizons: QIS Research and Innovation for
Nuclear Science.

\end{document}